# Trion Formation Hampers Single Quantum Dot Performance in Silane-Coated FAPbBr$_3$ Quantum Dots

Jessica Kline[1], Shaoni Kar[2], Benjamin F. Hammel[3], Yunping Huang[4], Zixu Huang[1], Seth R. Marder[3,4,5], Sadegh Yazdi[3,5], Gordana Dukovic[3,5,6], Bernard Wenger[7], Henry Snaith[2], David S. Ginger[1]*

[1]Department of Chemistry, University of Washington, Seattle, WA 98195, USA

[2] Department of Physics, University of Oxford, Clarendon Laboratory, Parks Road, Oxford, OX1 3PU, UK

[3]Materials Science and Engineering, University of Colorado Boulder, Boulder, CO 80309-0215, USA

[4]Department of Chemical and Biological Engineering, University of Colorado Boulder, Boulder, Colorado 80303, United States

[5]Renewable and Sustainable Energy Institute, University of Colorado Boulder, Boulder, CO 80309-0215, USA

[6]Department of Chemistry, University of Colorado Boulder, Boulder, CO 80309-0215, USA

[7]Helio Display Materials Limited, Wood Centre for Innovation, Quarry Rd, Headington, Oxford OX3 8SB, UK

* Corresponding author: dginger@uw.edu

**Abstract:** We explore silane-coated formamidinium lead bromide (FAPbBr$_3$) quantum dots as single photon emitters and compare them to FAPbBr$_3$ quantum dots passivated with a phosphoethylammonium derivative (PEAC$_8$C$_{12}$), which represents current state-of-the-art in zwitterionic molecular surface ligand passivation. We compare properties including single-photon purity ($g^{(2)}(\tau)$), linewidth, blinking, and photostability. We find that at room temperature, these silane-coated dots perform comparably to the PEAC$_8$C$_{12}$ passivation in terms of single-photon performance metrics, while exhibiting improvements in photostability. However, we find that at 4K, silane-coated FAPbBr$_3$ quantum dots perform worse than the PEAC$_8$C$_{12}$-passivated samples, exhibiting faster blue-shifting and photobleaching under illumination. Analysis of fluorescence lifetime intensity distributions from the photon-counting data indicates increased efficiency of fast non-radiative processes in the silane-coated quantum dots at 4K. We propose a trion-related degradation pathway at low temperatures that is consistent with the observed kinetics and estimate that at 4K with 6.1 μJ/cm$^2$, 472 nm excitation the silane-coated quantum dots build up double the trion population of their PEAC$_8$C$_{12}$-passivated counterparts.

Lead halide perovskite quantum dots are a promising material for a variety of optoelectronic applications, owing to their large absorption cross sections, high photoluminescence quantum yields,[1] narrow linewidths[2] and tunable emission.[3] These properties make them especially interesting for implementation in light emitting diodes (LEDs),[4] photovoltaics[5] and photodetectors.[6] Additionally perovskite quantum dots have recently emerged as a promising colloidal single photon source, given their strong anti-bunching behavior at all temperatures[7] and radiative lifetimes which approach the transform limit at low temperatures.[8] Indeed, Kaplan et al. have successfully demonstrated Hong-Ou-Mandel interference using colloidal perovskite quantum dots – proving this material can indeed serve as a source of indistinguishable single photons.[9] These advances are particularly exciting given emerging capabilities to pattern single colloidal emitters on demand, providing a pathway for chip-scale nanophotonic integration.[10] Despite these advantages, perovskite quantum dots still exhibit sub-optimal behaviors, including photoluminescence intermittency and spectral wandering (commonly known as blinking and spectral diffusion) that can impact long term photon indistinguishability. However, these are problems that are common to all types of quantum dots, indeed to most sources of single photons.[11]

In colloidal quantum dots, blinking and spectral diffusion depend on the quantum dot surface, and variations in the local environment.[11] As such, the first step to reducing blinking and spectral diffusion is to improve passivation of the quantum dot surface, which has the dual effect of filling traps and screening the quantum dot from the local environment. Colloidal quantum dot passivation generally falls into two standard motifs – the use of ligands that bind to the quantum dot surface or overgrown layers of a wider bandgap material (shells). For II-VI and III-V quantum dots, core-shell heterostructures have proven more successful in suppressing blinking and spectral diffusion than ligand passivation.[12] In fact, with the right shell researchers have been able to produce entirely non-blinking CdSe quantum dots.[13–15] This method has recently been extended to grow "colossal" shells[16] resulting in non-blinking quantum dots which are sufficiently large for deterministic positioning.[10] The success of this approach provides an excellent blueprint for creating a deterministically positionable single photon source, if a suitable shelling material can be identified.

Unfortunately, viable candidates for perovskite quantum dot core-shell heterostructures remain limited. As such perovskite quantum dot passivation has focused on ligand development, and many of the recent improvements in perovskite quantum dots as a single photon source have focused on optimizing the ligand chemistry.[17–21] Currently, state-of-the-art ligands for high performing single perovskite quantum dots are zwitterions derived from sulfobetaine and phosphoethlyammonium.[8,9,17] However, there is a significant amount of on-going work aimed at finding a good shelling material for perovskite quantum dots. Currently, oxide-based shells such as titania, alumina and silica are some of the most promising candidates.[22] Oxide-based shells provide significant improvements in the lifetimes of perovskite quantum dot thin films and increase the stability of thin films in the presence of heat, UV light, oxygen and water.[23–27] As such oxide-shelled perovskite quantum dots have received significant interest for LEDs.[26,28,29] These characteristics also suggest that oxide-based shells may work well for perovskite quantum dot single photon sources, as photostability at high fluences remains a challenge for this application as well.[30,31]

Here, we explore passivation of formamidinium lead bromide ($FAPbBr_3$) quantum dots with the diaminosilane N-(2-Aminoethyl)-3-aminopropyltriethoxysilane (AEAPTES), focusing on the performance of these quantum dots for single photon emission applications. To this end, we benchmark their performance against $FAPbBr_3$ quantum dots passivated with the current top-performing[17,32] ligand ($PEAC_8C_{12}$) both at room temperature and 4K. We find that silane-coated $FAPbBr_3$ quantum dots have excellent performance at room temperature but show reduced photostability at 4K when compared to the $PEAC_8C_{12}$-passivated dots.

**Results and Discussion**

We synthesized our PEAC$_8$C$_{12}$-passivated FAPbBr$_3$ quantum dots via the trioctylphosphine oxide/PbBr$_2$ method.[17,33] We first grew the dots using weakly bound ligands and exchanged them with the strongly binding PEAC$_8$C$_{12}$ post-synthesis using established protocols.[17] We synthesized the AEAPTES-modified FAPbBr$_3$ quantum dots via hot-injection[34] using oleylammonium and oleate as ligands and then ligand exchanged to AEAPTES as described in the method section. We chose AEAPTES for its ability to provide two functions: first, surface passivation *via* the head groups (amine and ammonium) and second, the possibility of cross linking by triethoxysilane hydrolysis. It is possible that AEAPTES may also react with FA$^+$ during the ligand exchange to form an imidazole-silane cation[35] which is functionally similar to mono-protonated AEAPTES (amine, ammonium and triethoxysilane). This first silane coating is essential for growing thicker silica shells using orthosilicates.[36] However, for these experiments, we did not overgrow the silane coating to accumulate significant thickness but compare the AEAPTES-exchanged quantum dots directly against PEAC$_8$C$_{12}$-exchanged dots. Figure S1 shows the chemical structures of PEAC$_8$C$_{12}$ and AEAPTES and Figure S2 shows that, under our synthesis, storage and sample preparation conditions, some AEAPTES undergoes triethoxysilane hydrolysis to form a partially cross-linked tail.

We carefully matched the quantum dot sizes (after ligand exchange/treatment) for this study to obtain similar absorption and emission spectra and minimize size-dependent effects. Figure 1 shows the solution characterization of the two near-identical batches of post-ligand exchange quantum dots. Both batches have photoluminescence quantum yields (PLQYs) near unity (Figures 1a and 1b), emission maxima of approximately 2.41 eV (Figures 1a and 1b) and photoluminescence lifetimes of approximately 4 ns (Figure 1c). Figure 1d shows the HAADF-STEM-measured size distributions for both samples and Figure S3 shows representative HAADF-STEM images. We find that the silane-coated quantum dots are ~1.1 nm larger than their PEAC$_8$C$_{12}$-passivated counterparts. However, given that the two samples have matched optical bandgaps, we ascribe this size difference to an orthosilicate monolayer (a ~ 0.5 nm) formed by hydrolysis of AEAPTES. Figure S4 shows that the silane-coated quantum dots also have a slightly larger ligand sphere (1.5 nm) than their PEAC$_8$C$_{12}$-passivated counterparts (1 nm). We also find that silane-coated quantum dots have a slightly broader size distribution – which is consistent with our observed difference in ensemble photoluminescence linewidth (121 vs 134 meV, Figures 1a and 1b).

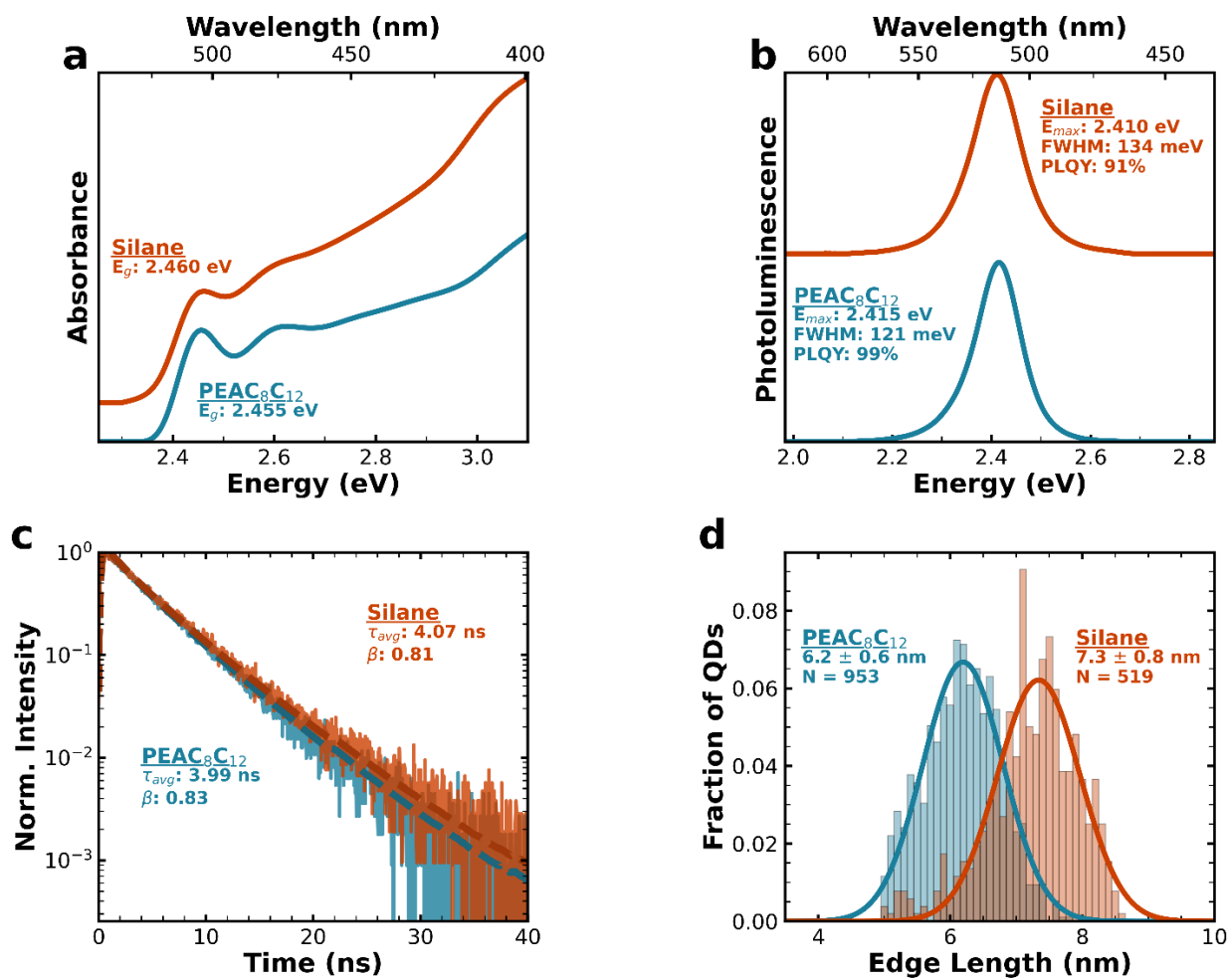

**Figure 1. Ensemble characterization of quantum dot samples**. a) absorbance spectra of PEAC$_8$C$_{12}$-passivated (blue) and silane-coated (orange) FAPbBr$_3$ quantum dots. Spectra are offset for clarity. PEAC$_8$C$_{12}$-passivated quantum dots have a first absorbance peak at 2.455 eV and silane-coated quantum dots have a first absorbance peak at 2.460 eV. b) photoluminescence spectra of PEAC$_8$C$_{12}$-passivated (blue) and silane-coated (orange) FAPbBr$_3$ quantum dots. Spectra are offset for clarity. PEAC$_8$C$_{12}$-passivated quantum dots have a photoluminescence peak is at 2.413 eV, a full width at half maximum of 121 meV and a PLQY of 99%. Silane-coated quantum dots have a photoluminescence peak is at 2.410 eV, has a full width at half maximum of 134 meV and a PLQY of 91%. c) photoluminescence lifetimes for our PEAC$_8$C$_{12}$-passivated (blue) and silane-coated (orange) FAPbBr$_3$ quantum dots. We fit the lifetimes to a stretched exponential (dashed lines, Equation S3) to best describe lifetime of a heterogeneous sample.[37] d) TEM-measured size distribution for our PEAC$_8$C$_{12}$-passivated (blue) and silane-coated (orange) FAPbBr$_3$ quantum dots. We fit the size distributions to a Gaussian (solid lines) to extract the average size. Figure S3 shows the HAADF-STEM images we used to extract the size distribution. Since the two samples have identical first absorption peaks and emission maxima, we ascribe the size difference between the two samples to the AEAPTES-passivation step.

We begin characterizing the behavior of our two samples as single quantum dots *via* widefield microscopy. Widefield microscopy is well-suited for characterizing the blinking behavior of a statistically significant number (N >> 100) of quantum dots in low-power regimes and can reveal disparate blinking behavior between comparable ensemble samples.[21] Figure S5 shows the blinking behavior we observed for our PEAC$_8$C$_{12}$-passivated and silane-coated quantum dots using widefield microscopy. We find that, at room temperature under low excitation power (405 nm, 9 mW/cm$^2$), both our PEAC$_8$C$_{12}$-passivated and silane-coated quantum dots perform well relative to current literature on room-temperature perovskite quantum dot blinking. Consistent with other reports,[17] we find that PEAC$_8$C$_{12}$-passivated quantum dots have an average ON% of 96% and a non-blinking fraction of 81%. We observed slightly improved performance from silane-coated quantum dots which have an average ON% of 95% and a non-blinking fraction of 90%. Even this minor improvement in performance is notable since PEAC$_8$C$_{12}$ is currently reported to be one of the best-performing surface passivation strategies for single-photon emitting perovskite quantum dots.[17,32]

Having observed encouraging blinking performance from both the PEAC$_8$C$_{12}$-passivated and silane-coated quantum dots under low excitation powers, we continue our room temperature single quantum dot characterization in a higher excitation power regime (472 nm, 6.1 µJ/cm$^2$) using pulsed confocal illumination at a repetition rate of 15.6 MHz. For comparison, this combination of pulse energy, duration, and frequency represents a time-averaged power of 95 W/cm$^2$ and a peak power of 3.1 MW/cm$^2$ which is orders of magnitude higher than typical test conditions for down-conversion phosphors.[38,39] Figures S6 and S7 show the full characterization of representative single quantum dots for PEAC$_8$C$_{12}$ passivation and silane coating respectively. We consider three main metrics to assess how well quantum dots perform as single photon emitters – linewidth, second order photon autocorrelation functions ($g^{(2)}(\tau)$) and weighted ON%. The ideal single photon emitters should have a narrow (transform-limited) linewidth, a low $g^{(2)}(\tau = 0)$ and a high ON%. Figure 2a shows the distribution of linewidths measured from our samples at room temperature. Single PEAC$_8$C$_{12}$-passivated quantum dots have a linewidth of 72 ± 2 meV (median ± interquartile deviation) and silane-coated quantum dots have a linewidth of 70 ± 3 meV. While the siliane-coated dots have a broader ensemble linewidths, the comparable single quantum dot linewidths indicate that much of the ensemble linewidth difference is due to differences in the size distribution (Figure 1d) – which is further supported by the difference in the distribution of the individual dot emission maxima (Figure S8).

Next, we compare the second order autocorrelation functions, $g^{(2)}(\tau)$, for silane-coated and PEAC$_8$C$_{12}$-passivated quantum dots. Figure 2b shows the distribution of measured values for $g^{(2)}(\tau = 0)$ which measures the single-photon purity of each quantum dot and confirms that we have measured a single quantum dot. We limit our sample size to quantum dots with a $g^{(2)}(\tau = 0)$ under 0.5, as that is the statistical limit for a photon source to be considered a single emitter. However, an excellent single-photon emitter should generally have a $g^{(2)}(\tau = 0)$ of less than 0.1.[7,31,40] By this metric, PEAC$_8$C$_{12}$-passivated and silane-coated quantum dots perform similarly at room temperature, with PEAC$_8$C$_{12}$-passivated quantum dots showing a slightly higher percentage of quantum dots with a $g^{(2)}(\tau = 0)$ under 0.1 (42 ± 10% vs 36 ± 10%), though the difference is close to the statistical uncertainty after measuring N = 53 and N = 65 quantum dots.

Finally, we compare the blinking behavior of our single quantum dots. Figure 2c shows the distribution of intensity-weighted ON percentages calculated from Change Point Analysis (CPA)-classified[41,42] blinking traces for both quantum dot samples. The intensity weighted ON percentage is defined in Equation S7, but in brief we deviate from traditional ON/OFF[17,43] and ON/OFF/GREY[44,45] analysis and treat all mid-intensity CPA-identified states as a linear combination of the ON (maximum intensity) and OFF ($I_{dark} + 3\sigma_{dark}$) states. This choice reduces the complex multi-level blinking dynamics of perovskite quantum dots ($n_{levels} > 10$)[46] to a simple system without completely ignoring the dynamics of the GREY states. We find that under our measurement conditions both silane-coated and PEAC$_8$C$_{12}$-passivated quantum dots perform similarly with average weighed ON percentages of 49 ± 13% and 46 ± 12% respectively.

Our choice to use an intensity-weighted ON percentage means that both blinking events and photobleaching can affect this figure of merit. Unfortunately, we see evidence of photobleaching during our measurement (Figures S6 and S7) which is unsurprising as perovskite quantum dots are known to photobleach at high excitation densities.[30,31] To distinguish photobleaching from blinking, we use large (10 s) time bins[47] and record the highest intensity observed in each window. We fit our maximum intensity traces to a linear decay where the slope is the average photobleaching rate. Figure S8 shows the distribution of the total intensity losses for both samples and Figure 2d shows, on average, how the maximum intensity changes in time for both samples. We find that, on average, PEAC$_8$C$_{12}$-passivated quantum dots lose 40% of their maximum intensity (approximately 6.7 kcps) during the 600 second measurement, under these rather intense excitation conditions. In contrast, silane-coated quantum dots are much more photostable and only lose 5-10% of their maximum intensity (approximately 1.8 kcps) over the same duration. These results indicate that at room temperature, silane coating results in significantly more photostable quantum dots than PEAC$_8$C$_{12}$ passivation. Since the silane coating is relatively thin, we speculate that in addition to providing a physical barrier, triethoxysilane polymerization and physical confinement may also serve to suppress photoinduced ligand desorption, thus contributing to increased stability.

Overall, using the state-of-the-art PEAC$_8$C$_{12}$ ligand as a benchmark, we find that silane-coated FAPbBr$_3$ quantum dots perform well as colloidal single photon emitters at room temperature. They have a comparable linewidth and weighted ON% to the PEAC$_8$C$_{12}$-passivated quantum dots. And while silane-coated quantum dots have a slightly worse $g^{(2)}(\tau = 0)$ distribution than PEAC$_8$C$_{12}$-passivated quantum dots, they are four times more photostable. The performance of our silane-coated quantum dots is even more impressive when we consider the size of our quantum dots (~ 6 nm edge length) and our excitation density (6.1 µJ/cm$^2$). These conditions should significantly increase the sensitivity of the observed single quantum dot performance to surface and environmental effects, as these effects are amplified in small quantum dots, causing small changes in passivation to result in large changes in performance.[48] And at the same time high excitation densities increase the impact of any detrimental higher order processes.[49,50] Given these conditions, it is all the more impressive that our silane-coated quantum dots keep up with, and even outperform, our PEAC$_8$C$_{12}$-passivated quantum dots; indicating that silane-coated perovskite quantum dots are in fact a promising direction to explore for room temperature single photon emitters.

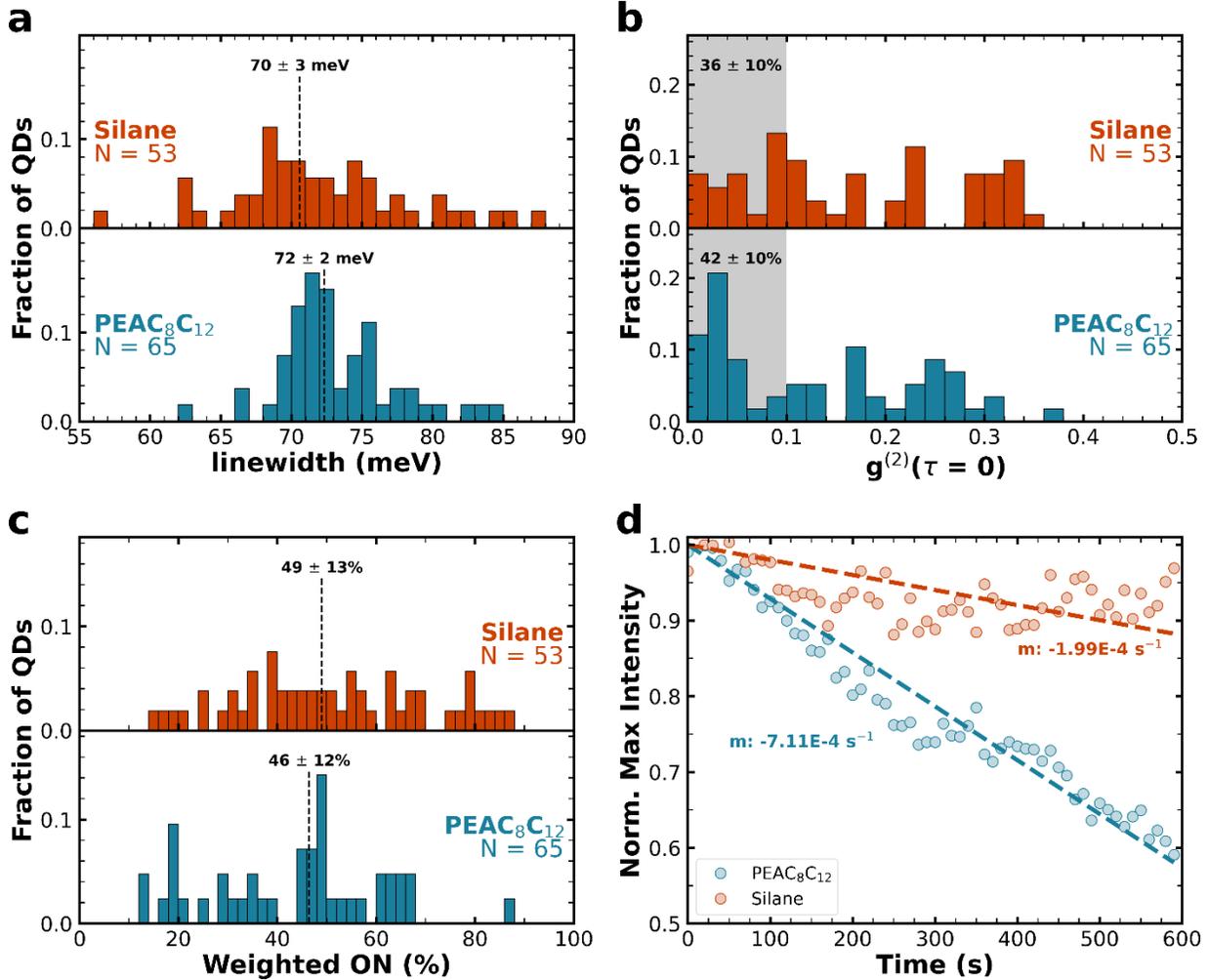

**Figure 2. Single quantum dot characterization at room temperature a)** Distribution of single quantum dot photoluminescence linewidths. Dashed lines and corresponding numbers represent the median ± interquartile deviation **b)** Distribution of $g^{(2)}(t=0)$ for single quantum dots. $g^{(2)}(t=0)$ values under 0.1 are considered excellent (grey box) and make up 42 ± 10% and 36 ± 10% of the measured quantum dots for $PEAC_8C_{12}$ passivation and silane coating respectively. **c)** Distribution of weighted ON%s for single quantum dots. Dashed lines and corresponding numbers represent the median ± interquartile deviation. **d)** Quantum dot photobleaching seen through a decreasing maximum intensity in time. Scattered points are the average of all ($N_{PEAC8C12}$ = 65, $N_{silane}$ = 53) single quantum dot photobleaching traces. Dashed lines are linear fits to the data of the form y = mx + 1.

Having explored the potential of our two samples as single photon sources at room temperature we continue our characterization at 4K. Figures S9 and S10 show the full 4K characterization of representative single quantum dots with $PEAC_8C_{12}$ passivation and silane coating, respectively. Our three main metrics to assess how well these quantum dots perform as single photon emitters remain the same as for the room temperature measurements. Additionally, at low temperatures spectral diffusion slows down, and linewidths narrow allowing us to assess the impact of spectral diffusion by measuring linewidths at both short (1s) and long (600 s) integration times. Figure 3a compares the average linewidth at a short integrations time (1 s) to the linewidth at a long integration time (600 s). While the $PEAC_8C_{12}$-passivated and silane-coated samples have similar linewidths at short times (20 ± 4 meV and 25 ± 5 meV respectively, Figure S11) silane-coated quantum dots have a much larger linewidth at long integration times (31 ± 8 meV and 44 ± 15 meV respectively, Figure S11). Surprisingly, this difference in short- and long-time linewidths does not appear to arise primarily from the random spectral fluctuations associated with spectral diffusion. Instead, as shown in Figure S12, the changing linewidth is strongly correlated to a continuous spectral blueshift. Figure 3b shows the average blue shift in time for both samples. While PEAC$_8$C$_{12}$-passivated quantum dots show a small spectral blue shift (averaging 8 meV over 20 minutes), silane-coated quantum dots blue shift nearly three times as far (averaging 23 meV over 20 minutes). Spectral blue-shifts are a common sign of photodegradation[51] – which is further confirmed by the observation of photobleaching during the blinking measurements (Figure S12).

Next, we compare the second order correlation functions $g^{(2)}(\tau)$ for silane-coated and PEAC$_8$C$_{12}$-passivated quantum dots. Figure 3c shows the distribution of measured values for $g^{(2)}(\tau = 0)$. Again, PEAC$_8$C$_{12}$-passivated quantum dots perform better than their silane-coated counterparts. 61 ± 10% of the PEAC$_8$C$_{12}$-passivated quantum dots have a $g^{(2)}(\tau = 0)$ under 0.1, compared to 49 ± 11% of the silane-coated quantum dots. The difference in the distribution of the $g^{(2)}(\tau = 0)$ values indicates that silane-coated quantum dots likely have either a larger chance of biexciton formation, or a higher biexciton quantum yield.

Finally, we compare the photoluminescence lifetimes and blinking behavior of our quantum dots. Figure 3d shows the distribution of intensity weighted ON percentages calculated from CPA classified blinking traces for both quantum dot samples. Once again, PEAC$_8$C$_{12}$-passivated quantum dots outperform their silane-coated counterparts. PEAC$_8$C$_{12}$-passivated quantum dots have an ON% of 66 ± 11%, while silane-coated quantum dots have an ON% of 55 ± 11%. We also compare the distribution of single quantum dot photoluminescence lifetimes, which are shown in Figure S13. We fit the lifetimes using a stretched exponential because the lifetime of a single quantum dot can vary during blinking events[52,53] – resulting in a measured lifetime analogous to a lifetime from a heterogenous sample.[37] Consistent with the observed higher ON%, PEAC$_8$C$_{12}$-passivated quantum dots also have longer lifetimes (383 ± 101 ps) than their silane-coated counterparts (276 ± 61 ps).

Overall, our characterization at 4K paints a different picture of silane-coated quantum dots as single photon emitters than the room temperature measurements. Silane-coated FAPbBr$_3$ quantum dots exhibit a persistent spectral blue shift, stronger photobleaching, a lower weighted ON% and shorter photoluminescence lifetimes than PEAC$_8$C$_{12}$-passivated quantum dots. These measurements indicate that at 4K PEAC$_8$C$_{12}$ provides better passivation for FAPbBr$_3$ quantum dots than the current generation of silane coating. This difference is especially puzzling given the favorable room temperature performance of these silane-coated perovskite quantum dots. Next, we explore the reasons for this change in performance at low temperature.

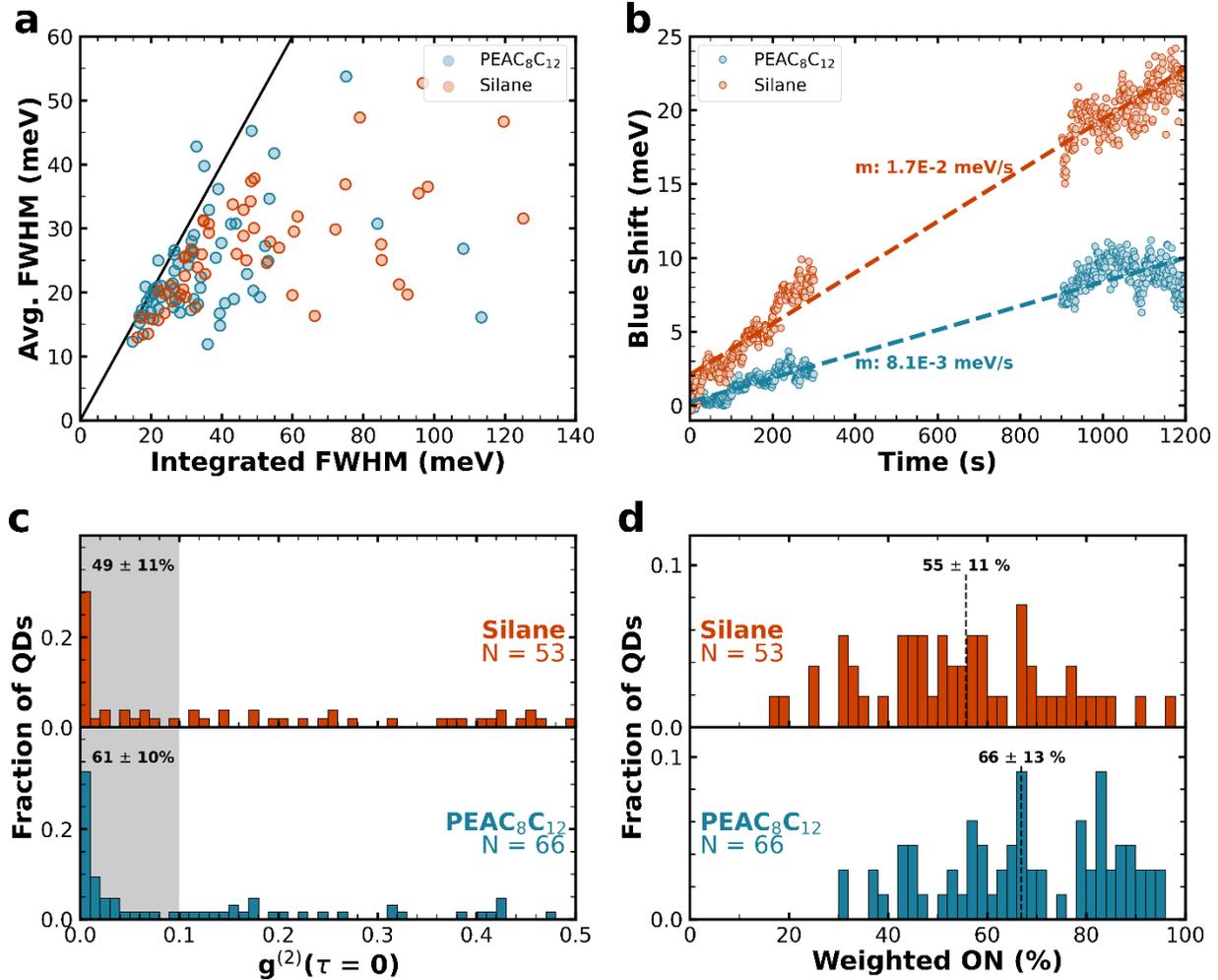

**Figure 3. Single quantum dot characterization at 4K a)** Average linewidth at short integration times (1s) compared to the linewidth at long integration times (600 s) for single quantum dots. y=x is represented by the solid black line. **b)** Photodegradation seen through a spectral blue shift in time. Dashed lines are linear fits to the data of the form y = mx + b. The gap from 300 to 900 seconds corresponds to the measurement time for quantum dot blinking, lifetime and $g^2(\tau)$. **c)** Distribution of $g^{(2)}(\tau = 0)$ for single quantum dots. $g^{(2)}(\tau = 0)$ values under 0.1 are considered excellent (grey box) and make up 61 ± 10% and 49 ± 11% of the measured quantum dots for PEAC$_8$C$_{12}$ passivation and silane coating respectively. **d)** Distribution of weighted ON%s for single quantum dots. Dashed lines and corresponding numbers represent the median ± interquartile deviation.

We find the first hint towards the cause of the decreased performance by exploring the underlying non-radiative decay mechanisms responsible for blinking at room temperature and 4K. Different non-radiative decay mechanisms have unique lifetime-intensity correlations, which can be easily identified by fluorescence lifetime intensity distributions (FLIDs). There are three non-radiative pathways which are most commonly considered responsible for quantum dot blinking – band-carrier (BC) trapping, Auger-Meitner recombination and hot-carrier (HC) trapping.[53,54] BC trapping is associated with the multiple recombination center model of quantum dot blinking, where the quantum dot remains neutral and blinking occurs because of short-lived traps.[53–55] In this model, blinking occurs as the non-radiative recombination rate changes while the radiative recombination rate remains constant resulting in a linear lifetime-intensity correlation.[53,54] Auger-Meitner mediated blinking is associated with quantum dot charging and trion formation, which could occur either by photoionization or long-lived traps.[53,54] In this scenario both the radiative and non-radiative recombination rates vary as the quantum dot blinks resulting in a non-linear correlation between the lifetime and intensity.[52–54] HC trapping is also associated with neutral quantum dots, but in this model carriers are trapped before relaxing to the band-edge resulting in a lifetime which is independent of intensity.[54] Importantly, these pathways are not mutually exclusive and a single quantum dot can show a combination of HC trapping, BC trapping and Auger-Meitner recombination.[52]

Representative FLIDs for $PEAC_8C_{12}$-passivated and silane-coated quantum dots at room temperature are shown in Figures 4a and 4b respectively. These FLIDs are constructed by summing the lifetime and intensity normalized single quantum dot FLIDs for each sample. We do not see the signature of HC trapping in any of our FLIDs at room temperature (unsurprising given that our excitation is only 200 meV above the bandgap) and as such discard HC trapping as a possible blinking mechanism in our samples under these measurement conditions. However, our FLIDs do show signs of both BC trapping and Auger-Meitner recombination. At room temperature, both samples show a predominantly linear relationship between lifetime and intensity indicating that BC trapping is the dominant non-radiative recombination channel. For a more detailed analysis we use Equation S8, which assumes that the overall FLID pattern is a linear sum of the two underlying patterns,[56] to resolve the relative percentage of non-radiative decay which can be attributed to BC trapping and Auger-Meitner recombination. The solid white line shows the fit of the representative FLIDs to Equation S8, and dashed lines represent the 100% BC and 100% Auger-Meitner mediated extremes. At room temperature blinking in $PEAC_8C_{12}$-passivated quantum dots is primarily mediated by BC trapping (99%), with less than 1% of blinking attributed to Auger-Meitner recombination. Blinking in silane-coated quantum dots is also primarily mediated by BC trapping (93%), but we find that Auger-Meitner recombination plays a more significant role at 7%. Figure S8 shows the distributions of BC trapping and Auger-Meitner mediated blinking for individual quantum dots at room temperature. These FLID patterns indicate that at room temperature non-radiative recombination predominately occurs *via* short-lived traps. However, silane-coated quantum dots likely have a slightly higher probability of trion and/or biexciton formation consistent with their observed tendency for larger $g^{(2)}(\tau = 0)$ values (Figure 2b).

Figures 4c and 4d show the representative FLIDs for $PEAC_8C_{12}$-passivated and silane-coated quantum dots acquired at 4K. At 4K, neither of our samples show a predominantly linear relationship between lifetime and intensity. Instead, both samples show a strong non-linear correlation indicating that Auger-Meitner recombination plays a more significant role in non-radiative recombination at 4K. To resolve the relative percentage of non-radiative decay which can be attributed to BC trapping and Auger-Meitner recombination we fit our FLIDs to Equation S8. At 4K we find that blinking in $PEAC_8C_{12}$-passivated quantum dots is primarily mediated by Auger-Meitner recombination (58%), although BC trapping still plays a significant role (42%). We find that blinking in silane-coated quantum dots is significantly more likely to be mediated by Auger-Meitner recombination (80%) with BC trapping playing a smaller role (20%). Figure S13 shows the distributions of BC trapping and Auger-Meitner mediated blinking for individual quantum dots at 4K.

The general trend of increased Auger-Meitner recombination contributions indicates that at 4K perovskite quantum dots tend to have long-lived traps while at room temperature the traps are mostly short-lived. And while the non-radiative decay mechanisms for silane-coated and $PEAC_8C_{12}$-passivated quantum dots are very similar at room temperature, at 4K Auger-Meitner recombination is significantly more likely in silane-coated quantum dots than in $PEAC_8C_{12}$-passivated quantum dots. This suggests that silane-coated quantum dots have an additional source of traps which play a significant role at 4K. As Auger-Meitner recombination is most likely to be associated with trion and biexciton formation – we further explore trion and biexciton formation in these samples to understand their differences in 4K performance.

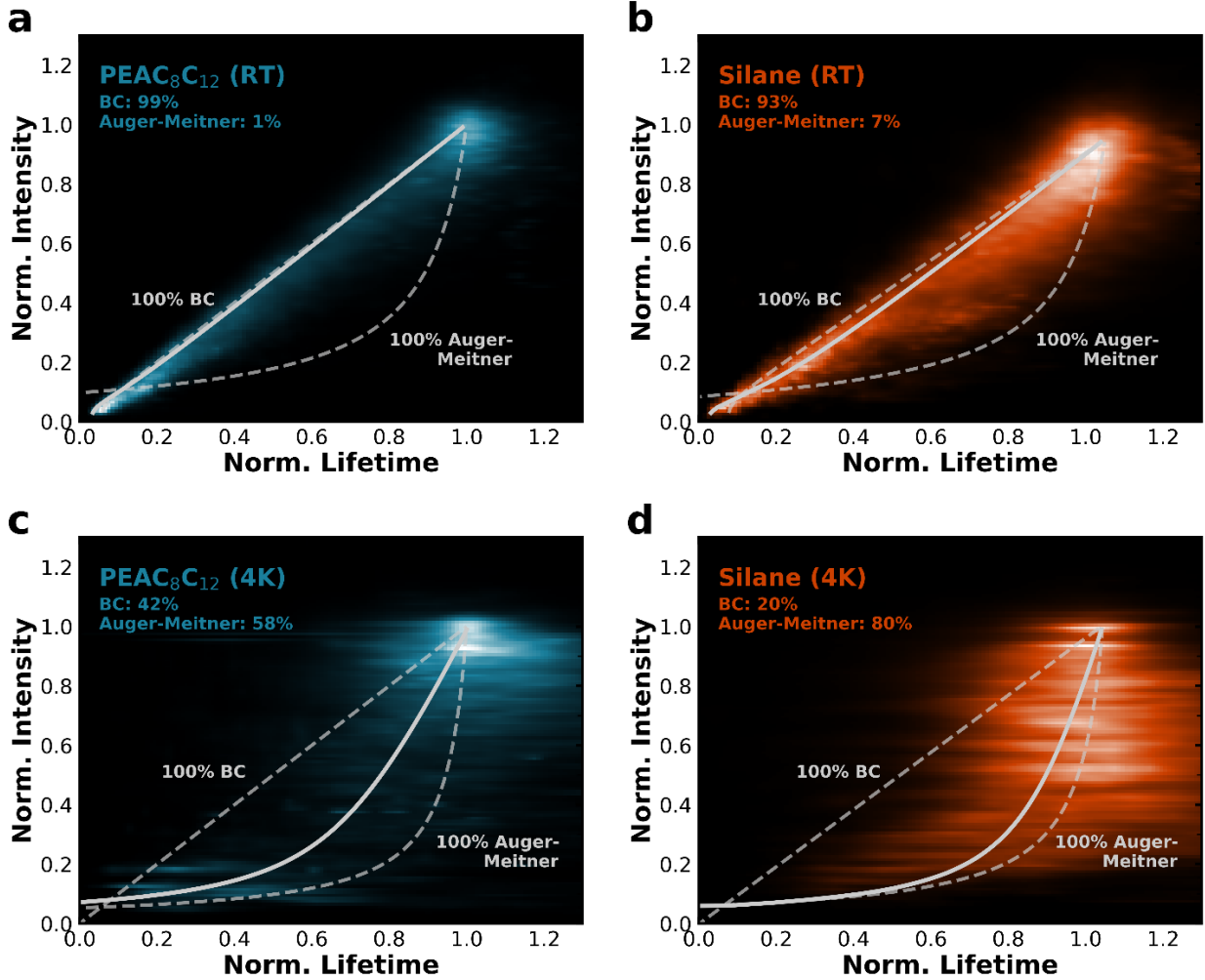

**Figure 4. Comparing non-radiative decay mechanisms at room temperature and 4K with Fluorescence Lifetime Intensity Distributions.** FLID plots summed from all ($N_{PEAC8C12}$ = 65, $N_{silane}$ = 53) single quantum dot to represent the average single quantum dot at room temperature for **a)** PEAC$_8$C$_{12}$-passivated and **b)** silane-coated. FLID plots aggregated from single quantum dots ($N_{PEAC8C12}$ = 66, $N_{silane}$ = 53) to represent the average single quantum dot at 4K for **c)** PEAC$_8$C$_{12}$-passivated and **d)** silane-coated. The solid line represents the best fit of the data as a combination of BC trapping and Auger-Meitner recombination, while the dashed lines represent the two extremes. While the non-radiative decay mechanisms for silane-coated and PEAC$_8$C$_{12}$-passivated quantum dots are very similar at room temperature, at 4K Auger-Meitner recombination is significantly more likely in silane-coated quantum dots than in PEAC$_8$C$_{12}$-passivated quantum dots. This suggests that silane-coated quantum dots have an additional source of traps which play a significant role at 4K.

First, we look for signs of trion and biexciton formation in the fluence dependent PLQY. Figure 5a describes a simple possible physical model for the formation of trions and biexcitons in quantum dots. In this model we consider three states: the exciton (X), the trion (X$^\pm$) and the biexciton (XX). The formation rate of the exciton is linear with respect to the excitation density, while the trion and biexciton formation rates have a second-order dependence on the excitation density. Each state has an associated radiative ($k_{r,n}$) and non-radiative ($k_{nr,n}$) recombination rate and the biexciton has an additional trapping rate ($k_t$) associated with the long-term trapping of a charge carrier which populates the trion.

Figure 5b shows the fluence dependent PLQY of quantum dot thin films made from both samples. We calculate the expected number of excitons formed per pulse (<N>) by estimating the absorption cross section from the fluence dependent intensity of the thin film (Equation S11 and Figure S14).[57,58] The highest <N> value corresponds to our single quantum dot characterization fluence (6.1 µJ/cm$^2$). Qualitatively, the PLQY of the silane-coated quantum dots has a steeper roll off than the PLQY of the PEAC$_8$C$_{12}$-passivated quantum dots which is consistent with increased higher-order non-radiative recombination. To check that the observed fluence dependence is consistent with the trion interpretation from the FLID distributions, we next fit the fluence dependent PLQY using Equations 1a-1e which are derived from the physical model depicted in Figure 5a:

$$QY(\langle N \rangle) = \frac{N_X(\langle N \rangle)k_{r,X} + N_{X\pm}(\langle N \rangle)k_{r,X\pm} + N_{XX}(\langle N \rangle)k_{r,XX}}{N_X(\langle N \rangle)(k_{r,X}+k_{nr,X}) + N_{X\pm}(\langle N \rangle)(k_{r,X\pm}+k_{nr,X\pm}) + N_{XX}(\langle N \rangle)(k_{r,XX}+k_{nr,XX}+k_t)} \quad (1a)$$

$$\frac{\partial N_X}{\partial t} = -N_X(k_{r,X} + k_{nr,X}) + N_{XX}(k_{r,XX} + k_{nr,XX}), \qquad N_X(0) = n \quad (1b)$$

$$\frac{\partial N_{X\pm}}{\partial t} = N_{XX}(k_t) - N_{X\pm}(k_{r,X\pm} + k_{nr,X\pm}), \qquad N_{X\pm}(0) = 0 \quad (1c)$$

$$\frac{\partial N_{XX}}{\partial t} = -N_{XX}(k_{r,XX} + k_{nr,XX} + k_t), \qquad N_{XX}(0) = n^2 \quad (1d)$$

$$\langle N \rangle = n + 0.5n^2 \quad (1e)$$

Equation 1a describes the expected behavior of the PLQY with changing excitation density; and depends on the fluence dependent state populations $N_n(\langle N \rangle)$ and the various radiative and non-radiative recombination rates. Equations 1b-1d are a system of differential equations and initial conditions which describe how the population of each state changes in time. Equation 1e describes the expected branching ratio between exciton and biexciton generation for a given excitation density based on Poisson statistics. The system of equations in Equations 1b-1d describe an initial value problem which can be solved for the fluence dependent state populations. To find physically reasonable recombination rates we constrain our radiative and non-radiative rates based on the lifetime and quantum yield of the samples at our lowest fluence (0.45 nJ/cm$^2$, Figure S15) and the expected statistical scaling of recombination rates in higher order states.[59] These constraints, along with model limitations and alternative PLQY models are discussed in more detail in the Supplementary Information under the heading "Fluence Dependent PLQY" and in Figure S16.

Table 1 shows the recombination rates found by fitting the data in Figure 5b using Equations 1a-1e. Additionally, as shown in Table S1 and Figure S15, the recombination rates found from fitting the fluence dependent PLQY data are consistent with the experimental dynamics of the exciton and trion states from our single quantum dot data and our experimental fluence dependent lifetimes. The fitting reveals two primary differences in the recombination dynamics of the PEAC$_8$C$_{12}$-passivated and silane-coated quantum dots. First, silane-coated quantum dots have larger biexciton and trion non-radiative rate constants leading to lower state quantum yields, consistent with the low temperature FLIDs (Figures 4c and 4d) where silane-coated quantum dots are more likely to recombine *via* Auger-Meitner recombination. And second, silane-coated quantum dots have a larger trapping rate constant ($k_t$), suggesting that silane-coated quantum dots also have a larger population of trions due to a more efficient trapping process.

**Table 1.** Recombination rates and quantum yields for exciton (X), trion (X$^\pm$) and biexciton (XX) states extracted from fluence dependent PLQY measurements.

| | Exciton (X) | | | Trion (X$^\pm$) | | | Biexciton (XX) | | | | |
|---|---|---|---|---|---|---|---|---|---|---|---|
| | $k_{r,X}$ (ns$^{-1}$) | $k_{nr,X}$ (ns$^{-1}$) | $QY_X$ | $k_{r,X\pm}$ (ns$^{-1}$) | $k_{nr,X\pm}$ (ns$^{-1}$) | $QY_{X\pm}$ | $k_{r,XX}$ (ns$^{-1}$) | $k_{nr,XX}$ (ns$^{-1}$) | $QY_{XX}$ | $k_t$ (ns$^{-1}$) | $P_{t,XX}$ |
| **PEAC$_8$C$_{12}$** | 1.9 ± 0.2 | 0.06 ± 0.01 | 0.97 | 2.8 ± 0.3 | 0.12 ± 0.02 | 0.96 | 6.5 ± 0.7 | 0.46 ± 0.05 | 0.68 | 2.7 ± 0.3 | 0.28 |
| **Silane** | 1.9 ± 0.2 | 0.11 ± 0.01 | 0.95 | 2.9 ± 0.4 | 0.81 ± 0.07 | 0.78 | 6.7 ± 0.8 | 3.0 ± 0.4 | 0.39 | 7.6 ± 0.8 | 0.44 |

Having shown that silane-coated quantum dots have a steeper PLQY roll-off primarily due to increased Auger-Meitner recombination, we look to further assess the relative probability of forming a trion in both of our samples. To do this we look for the signature of trion emission in our single quantum dot photoluminescence spectra. At 4K single quantum dots are expected to have five spectral contributions – the zero-phonon line (ZPL), trion emission, biexciton emission and two longitudinal optical (LO) phonons. The other spectral contributors are expected to be red-shifted from the ZPL, although emission energy and intensity are size-dependent. The emission contributions of the LO phonons are size-independent and occur at -4.9 and -19.5 meV,[60,61] while the trion and biexciton emission contributions are size-dependent.[50,60,62] For FAPbBr$_3$ quantum dots with an edge-length between 5.5 and 6.5 nm, the trion and biexciton emission peaks are expected to be between -30 and -20 meV and -50 and -40 meV respectively.[60]

Since our samples exhibit significant spectral variations in time, we chose to correct our spectra according to Gumbsheimer et al.[63] in order to better resolve the emission structure. After correcting our spectra, we integrate across the collection time and look for the signatures of trion and biexciton emission. Figure 5c shows representative corrected and integrated single quantum dot photoluminescence spectrum for both samples.

Qualitatively, the spectra for silane-coated and PEAC$_8$C$_{12}$-passivated quantum dots show one major difference: silane-coated quantum dots have a pronounced shoulder around -20 meV. This is exactly the range of emission which should correspond to trion emission in approximately 6 nm FAPbBr$_3$ quantum dots.[60] This peak is likely broadened due to overlap with the LO phonon mode at -19.6 meV, leading to a broader flat shoulder instead of a distinct peak. Both samples have minimal emission in the range of biexciton emission (-50 to -40 meV) which is consistent with our measured in $g^{(2)}(\tau = 0)$ values (Figure 3b). The dashed lines in Figure 5c show fits to the spectra using a five Gaussian model which considers emissive contributions from the zero phonon line, the two LO phonons and trion and biexciton emission. Figure S17 shows the details of this fitting for the representative spectra shown in Figure 5c and Figure S18 shows the distribution of emission by state extracted from this fitting for all of our quantum dots. On average we find that the PEAC$_8$C$_{12}$-passivated quantum dots have emission split 80/16/4 between the exciton, trion and biexciton respectively, while silane-coated quantum dots have emission split 72/22/6 between the exciton, trion and biexciton.

Using the results of these fittings we can estimate the relative population of each state by correcting for quantum yields, trapping and fluence-dependent emission intensity trends according to Equation 2a:

$$I_n = N_n I_{ex}^{\beta_n} QY_n \quad (2a)$$

$$N_{eff,n} = (1 - P_{t,n})N_n \quad (2b)$$

Where $I_n$ is the measured emission intensity of a given state, $N_n$ is the population of a given state, $I_{ex}$ is the excitation intensity, $\beta_n$ describes the fluence-dependent emission intensity trend of a given state and $QY_n$ is the quantum yield of a given state. For states which have an additional trapping pathway (biexciton) we further calculate the population that is not trapped ($N_{eff,n}$) based on the probability of trapping ($P_{t,n}$) using Equation 2b. Using the experimental state quantum yields and trapping probabilities calculated from our fluence dependent PLQY model (Table 1) and literature fluence-dependent emission intensity trends[50,60] we calculate the probability of occupying each state, which is shown in Table 2. We find that, under our measurement conditions, the trion state of silane-coated quantum dots is twice as likely to be occupied as its PEAC$_8$C$_{12}$-passivated counterpart.

**Table 2.** Relative population of exciton, trion and biexciton states determined from single quantum dot photoluminescence spectra

|  | **PEAC$_8$C$_{12}$** | **Silane** |
|---|---|---|
| **Exciton** | 0.92 ± 0.08 | 0.82 ± 0.03 |
| **Trion** | 0.08 ± 0.04 | 0.16 ± 0.02 |
| **Biexciton** | 0.01 ± 0.01 | 0.02 ± 0.01 |

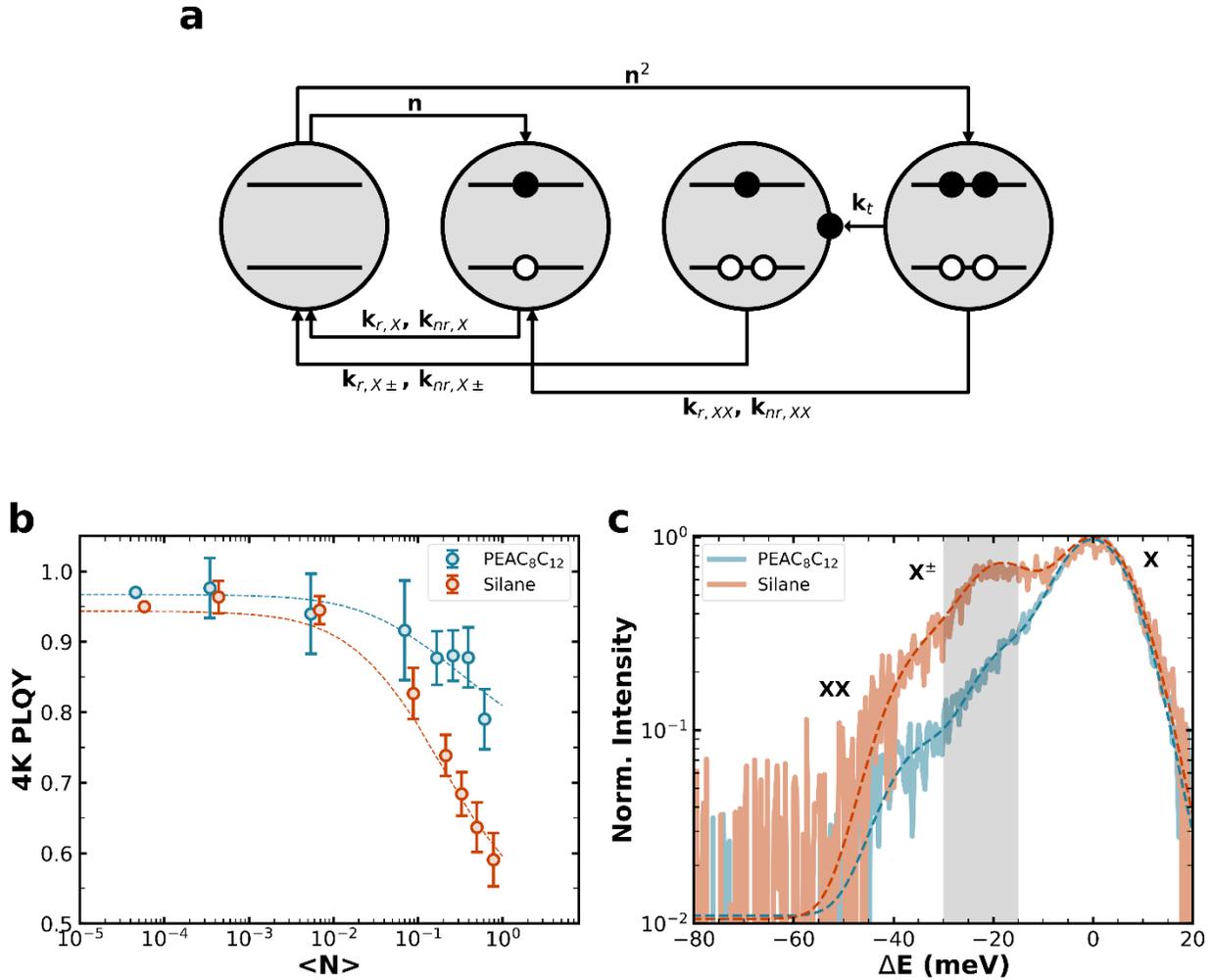

**Figure 5. Evidence for increased trion formation in silane-coated quantum dots at 4K. a)** Photophysical model for the formation and recombination of excitons, trions and biexcitons in quantum dots **b)** Fluence dependent PLQY measurements for quantum dot thin films. Dashed lines are the fit to Equation 1, see Table 1 for fit parameters. Silane-coated quantum dots see an early roll off due to higher order non-radiative recombination. **c)** Representative long integration time single quantum dot photoluminescence spectra aggregated from spectral diffusion corrected short integration time spectra (solid lines). Dashed lines are the fit to Equation S10. Silane-coated quantum dots show more emission from the trion (X$^\pm$) states.

Together, the observed fluence dependent PLQY trends, single quantum dot photoluminescence spectra and FLIDs paint a self-consistent picture where silane-coated quantum dots perform worse at 4K due to trion formation. We find that the trion state in silane-coated quantum dots has a lower quantum yield and higher population at high fluence than the trion state in its PEAC$_8$C$_{12}$-passivated counterpart. This combination of factors means that silane-coated quantum dots have significantly more Auger-Meitner non-radiative recombination under intense fluence at low T and increases the amount of time the quantum dot spends charged.

We hypothesize what could be two likely causes of long-lived traps and surface charging at 4K with our silane-coated quantum dots: head group variations and tail group variations. AEAPTES, our silane passivation, is a diamine– meaning that it has two amines which could act as head groups. To displace the as-synthesized oleylammonium passivation, it is likely that one of AEAPTES' amines becomes an ammonium during the ligand exchange providing strong passivation for A-site vacancies in the perovskite lattice. However, X-site vacancies can only be passivated by the remaining amine or residual oleate. Both oleates and amines are considered weak passivants and their usage for X-site passivation in these quantum dots may result in an additional source of traps with a higher occupation probability at 4K. In contrast PEAC$_8$C$_{12}$ is a zwitterionic ligand and as such can passivate both A- and X-site vacancies well. The effectiveness of using solely cationic or dicationic ligands for single quantum dot applications is still being debated, with both good[18,64] and bad[32] performances reported, but we consider this difference in head groups to be a potential source of the worse performance of our silane-coated quantum dots at 4K.

We propose that a second potential cause of long-lived traps and surface charging at 4K for our silane-coated quantum dots could be the silane tail. Silane layers, and their associated silanols and siloxanes are often considered a poor optoelectronic material due to deep carrier traps [65–67] and have proved largely ineffective in organic LEDs. Additionally, the blinking dynamics of silica-shelled CdSe/CdS quantum dots show evidence for the formation of a large number of trion, biexciton and other higher-order charged states at room temperature thanks to traps at the CdS-silica interface.[59,68] While AEAPTES does not directly create a perovskite-silane interface[69,70] for a similar trapping mechanism to silica-shelled CdSe/CdS quantum dots, the silane tail is plausibly within tunneling distance of the quantum dot surface which would allow for trapping within the silane tail, consistent with the larger trapping rate constant for silane-coated quantum dots.

**Conclusion**

Here we have explored the performance of silane-coated FAPbBr$_3$ quantum dots as single photon emitters. We find that at room temperature silane coating performs as well as passivation from the literature-best ligand PEAC$_8$C$_{12}$. Both passivation methods have narrow linewidths (~71 meV), low $g^{(2)}(\tau = 0)$ values (> 36% below 0.1) and a comparable ON% (50%). We also find that at room temperature silane-coated quantum dots are more photostable than their PEAC$_8$C$_{12}$-passivated counterparts. However, at 4K we find that silane-coated quantum dots perform worse than their PEAC$_8$C$_{12}$-passivated counter parts with broader linewidths, higher $g^{(2)}(\tau = 0)$ values and a lower ON%.

From FLIDs acquired at 4K we observe that silane-coated quantum dots tend toward Auger-Meitner dominated non-radiative recombination. This result suggests that silane-coated quantum dots form more trions and other higher order excitonic states such as biexcitons under high fluences at low temperature. This result is further supported by single particle spectra and fluence dependent PLQY measurements which indicate that the silane-coated quantum dots are two times more likely to form trions than the PEAC$_8$C$_{12}$-passivated quantum dots. We suggest that this increased trion formation, and the subsequent prevalence of surface charging, is the cause of the photodegradation we observe at 4K.

Although, our silane coating performs worse than PEAC$_8$C$_{12}$ passivation at 4K – its excellent performance at room temperature indicates that passivation using similar motifs are worth further exploration. One consideration is an aminoalkyl triethoxysilane with a longer alkyl chain which could prevent surface charges from tunneling to the silane tail or changing the amine/ammonium head group to a zwitterionic head group to provide better X-site passivation. Alternatively, we may be able to keep the benefits of passivation using a cross-linked ligand and remove the 4K trapping mechanism by choosing a ligand with a different cross-linking moiety, such as a thiol, which is less likely to have a large trap density.

**Methods**

Our synthesis, characterization and data analysis are described in detail in the supplementary information under the headings "Synthesis", "Characterization" and "Data Analysis". But in brief our PEAC$_8$C$_{12}$-passivated quantum dots were synthesized according to Morad et al. [17] Our silane-coated quantum dots were synthesized according to Protesecu et al.[71] and then ligand-exchanged to AEAPTES in a glovebox.

Single quantum dot films were prepared by diluting by a factor of 1,000 in toluene. Quantum dots were then spun coat onto a 1 mm quartz substrate (confocal measurements) or #1.5 coverslip (widefield measurements). Single quantum dot optical characterization was carried out on a home-built confocal microscope integrated with a cryostat (AttoDry800, Attocube). Samples were illuminated with a pulsed laser at 472 nm (15.647 MHz, 95 W/cm$^2$, 6.1 μJ/cm$^2$, NKT Photonics). Spectra were collected on an Isoplane 320 and Pixis 400 (Princeton Instruments) and two avalanche photodiodes (35 ps IRF, Micro Photon Devices) were used to collect time-tagged-time-resolved data. Analysis of the time-tagged-time-resolved data was performed according to Palstra et al.[41] Widefield microscopy characterization of single quantum dots was collected and analyzed according to Gallagher et al.[21]

**Supporting Information**

Additional experimental details, synthesis methods, and characterization.

**Author Contributions**

The manuscript was written through contributions of all authors. All authors have given approval to the final version of the manuscript.

**Acknowledgements**

This work, and the roles of J.K., B.F.H, Y.H., S.R.M., S.Y., G.D. and D.S.G, were primarily supported by the National Science Foundation under the STC IMOD Grant (No. DMR-2019444). B.F.H. acknowledges support from the National Science Foundation through the Graduate Research Fellowship Program (NSF-GRFP) under Grant No. DGE 2040434. S.K., H.S. and B.W. acknowledge funding from the European Union's Horizon 2020 research and innovation program under the Marie Skłodowska-Curie grant agreement No. 956270 for supporting synthesis and ligand exchange of the silane-coated quantum dots. Z.H. acknowledges gift funding from the Washington Research Foundation for supporting ATR-IR characterization of the quantum dots.

The authors acknowledge the use of facilities and instruments at the Photonics Research Center (PRC) at the Department of Chemistry, University of Washington, as well as that at the Research Training Testbed (RTT), part of the Washington Clean Energy Testbeds system. HAADF-STEM imaging was carried out at the Facility for Electron Microscopy of Materials at the University of Colorado Boulder (CU FEMM, RRID: SCR_019306).

J.K. would like to acknowledge Akash Dasgupta (University of Washington) for discussions regarding modeling fluence dependent PLQY in quantum dots.

The authors declare no competing financial interests.